\documentclass[letterpaper]{JHEP3}
\usepackage{amssymb}

\usepackage{epsfig}

\usepackage{amsfonts}
\usepackage{amsmath}

\newcommand{\beq}{\begin{equation}}
\newcommand{\eeq}{\end{equation}}
\newcommand{\beqs}{\begin{eqnarray}}
\newcommand{\eeqs}{\end{eqnarray}}

\newcommand{\tr}{\mbox{\rm tr}}
\newcommand{\ra}{\rightarrow}
\numberwithin{equation}{section}

\abstract{ We present arguments for the existence of new black string solutions 
with negative cosmological constant. These higher-dimensional 
configurations have no dependence on the `compact' extra dimension, and 
their conformal infinity is the product of time and $S^{d-3}\times R$ or 
$H^{d-3}\times R$. The configurations with an event horizon topology 
$S^{d-2}\times S^1$ have a nontrivial, globally regular limit with 
zero event horizon radius. We discuss the general properties of such 
solutions and, using a counterterm prescription, we compute their 
conserved charges and discuss their thermodynamics.
 Upon performing a dimensional reduction we prove that the reduced action 
has an effective $SL(2,R)$ symmetry. 
This symmetry is used to construct non-trivial 
solutions of the 
Einstein-Maxwell-Dilaton system with a Liouville-type potential for the dilaton in $(d-1)$-dimensions.}  
\preprint{hep-th/yymmddd}

\title{Black string solutions with negative cosmological constant}
\author{Robert B. Mann,$^{1,3}$\thanks{E-mail: \texttt{mann@sciborg.uwaterloo.ca}} \
Eugen Radu,$^2$\thanks{E-mail: \texttt{radu@thphys.nuim.ie}} \ Cristian Stelea$^{3}$\thanks{E-mail: \texttt{cistelea@uwaterloo.ca}} \\
$^{1}$Perimeter Institute for Theoretical Physics, Waterloo, Ontario N2L 2Y5, Canada\\
$^{2}$Department of Mathematical Physics,  National University of Ireland, Maynooth, Ireland\\
$^{3}$Department of Physics, University of Waterloo Waterloo, Ontario N2L 3G1, Canada}

\begin{document}
\section{Introduction}

Since its inception as a theory of gravity a tremendous amount of 
work has been done in General Relativity on the subject of black hole 
solutions of the Einstein field equations. As the end points of 
gravitational collapse, black holes are among the most interesting objects 
predicted to exist. The physics of black holes has quickly become one 
fascinating area of research as thermodynamics, gravity and quantum theory 
are intertwined in the black hole description. In recent years there 
has been a great deal of attention devoted towards research in  
black hole physics in four and higher dimensions. In more than four 
dimensions there are generically many available phases of black objects, 
with rich phase structures and interesting phase transitions between 
different kinds of black holes.

The physics of asymptotically Anti-de Sitter (AdS) black hole 
solutions is of particular interest  due to the AdS/CFT conjecture. The 
thermodynamic properties of black holes in AdS offers the 
possibility of studying the nonperturbative aspects of certain conformal field 
theories. For example, the Hawking-Page phase transition 
\cite{Hawking:1982dh} between the five dimensional spherically symmetric black 
holes and the thermal AdS background was interpreted by Witten, through 
AdS/CFT, as a thermal phase transition from a confining to a 
deconfining phase in the dual four dimensional  $\mathcal{N}=4$  super Yang-Mills 
(SYM)
theory \cite{Witten:1998zw}. Similarly, the phase structure of Reissner- 
Nordstr\"om-AdS black holes has a striking resemblance to that of van 
der Waals-Maxwell liquid-gas systems \cite{Chamblin:1999tk}. Also, in 
the presence of a negative cosmological constant $\Lambda$,  so-called 
topological black holes have been found to exist (see \cite{Mann:1997iz} 
for reviews of the subject). For such exotic black holes, the event 
horizon topology is no longer a sphere but can be any Einstein space 
with negative, positive or zero curvature. These results at 
least partially motivate attempts to find new black hole solutions with negative 
cosmological constant.

In this paper we present arguments for the existence of yet another 
class of configurations, which we interpret as the AdS counterparts of 
the $\Lambda=0$ uniform black string solutions \cite{bs}.  The black 
string solutions, present for $d\geq 5$ spacetime dimensions, exhibit 
new features that have no analogue in the black hole case. In the 
absence of a cosmological constant, the simplest vacuum static 
solution of this type is  found by assuming translational symmetry along the 
extra coordinate direction and taking the direct product of the 
Schwarzschild solution and the extra dimension. This corresponds to a 
uniform black string with horizon topology $S^{d-3}\times S^1$. Though 
this solution exists for all values of the mass, it is unstable below a 
critical value as first shown by Gregory and Laflamme 
\cite{Gregory:1993vy}. A branch of non-uniform black string solutions, depending 
also on the extra dimension was found in 
\cite{Gubser:2001ac,Wiseman:2002zc} (see also the recent work \cite{Kleihaus:2006ee}). The five 
dimensional AdS counterparts of such uniform black strings have been 
discussed in a more general context in a recent paper 
\cite{Copsey:2006br}. There are also known exact solutions for magnetically charged black strings in five-dimensional AdS backgrounds \cite{Chamseddine:1999xk}. Their properties have been further discussed in \cite{Klemm:2000nj} and they have been extended to higher dimensions in \cite{Sabra:2002xy}. However, for these solutions the magnetic charge of the black strings depends non-trivially on the cosmological constant and their limit (if any) in which the magnetic charge is sent to zero in order to recover the uncharged black strings in AdS is still unknown. Other interesting solutions whose boundary topology is a fibre bundle $S^1\times S^1\hookrightarrow S^2$ have been found in 
\cite{mann1} and later generalised to higher dimensions in \cite{Lu}.\footnote{Other locally asymptotically AdS geometries with non-trivial boundary geometries and topologies can be found in \cite{Astefanesei:2005eq}.}

Here we generalise the black string configurations of Ref. \cite{Copsey:2006br} to higher dimensions, 
finding also topological black string solutions with the $(d-3)$-dimensional sphere $S^{d-3}$ 
being replaced by a $(d{-}3)$--dimensional hyperbolic space. The 
solutions with  $S^{d-3}\times S^1$ topology of the event horizon have a 
nontrivial zero event horizon limit. We examine the general properties 
of these configurations solutions and compute their 
mass, tension and action by using a counterterm prescription. More generally, we can replace the $(d-3)$-dimensional angular sector with any Einstein space with positive/negative curvature leading to solutions with non-trivial boundary topologies.

By dimensionally reducing our black string solutions we find non-trivial black hole solutions of the Einstein-Dilaton system with a Liouville potential for the dilaton. We also prove that the reduced action has an effective $SL(2,R)$-symmetry. We use this symmetry to generate new charged solutions of the Einstein-Maxwell-Dilaton equations with a Liouville potential. We also compare our solutions with 
previously known charged black hole solutions. 

Our  paper is structured as follows: in the next section we explain 
the model and derive the basic equations, while in section $3$ we 
present a  computation of the physical quantities of the solutions such 
as mass, tension and action. The general properties of the solutions 
are presented in section $4$ where we show results obtained by numerical calculations. In the following sections we consider the dimensional reduction of our solutions to $(d-1)$-dimensions, proving the 
$SL(2,R)$-symmetry of the reduced action and using it explicitly to generate charged solutions of the Einstein-Maxwell-Dilaton system with Liouville potential. We give our conclusions and remarks in the final section.

\section{The model}
\subsection{Action principle and field equations}
We start with the following action principle in $d$-spacetime 
dimensions
\begin{eqnarray}
\label{action}
I_0=\frac{1}{16 \pi G}\int_{\mathcal{M}} d^d x \sqrt{-g}
 (R-2 \Lambda)
-\frac{1}{8 \pi G}\int_{\partial\mathcal{M}} d^{d-1} 
x\sqrt{-\gamma}K,
\end{eqnarray}
where  $\Lambda=-(d-1)(d-2)/(2 \ell^2)$ is the cosmological constant.

We consider the following parametrization of the $d$-dimensional line 
element (with $d \geq 5$)
\begin{eqnarray}
\label{metric} 
ds^2=a(r)dz^2+ \frac{dr^2}{f(r)}+r^2d\Sigma^2_{k,d-3}-b(r)dt^2
\end{eqnarray}
where the $(d{-}3)$--dimensional metric $d\Sigma^2_{k,d-3}$ is
\begin{equation}
d\Sigma^2_{k,d-3} =\left\{ \begin{array}{ll}
\vphantom{\sum_{i=1}^{d-3}}
 d\Omega^2_{d-3}& {\rm for}\; k = +1\\
\sum_{i=1}^{d-3} dx_i^2&{\rm for}\; k = 0 \\
\vphantom{\sum_{i=1}^{d-3}}
 d\Xi^2_{d-3} &{\rm for}\; k = -1\ ,
\end{array} \right.
\end{equation}
where $d\Omega^2_{d-3}$ is the unit metric on $S^{d-3}$. By $H^{d-3}$ 
we will understand the $(d{-}3)$--dimensional hyperbolic space, whose unit 
metric  $d\Xi^2_{d-3}$ can be obtained by analytic continuation of 
that on $S^{d-3}$. The direction $z$ is periodic with period $L$.

The Einstein equations with a negative cosmological constant imply 
that the metric functions $a(r)$, $b(r)$ and $f(r)$ are solutions of 
the following equations:
\begin{eqnarray}
\label{ep1} 
f'=\frac{2k(d-4)}{r}
+\frac{2(d-1)r}{\ell^2}
-\frac{2(d-4)f}{r}
-f\left(\frac{a'}{a}+\frac{b'}{b} \right)~,
\end{eqnarray}
\begin{eqnarray}
\label{ep2} 
b''&=&\frac{(d-3)(d-4)b}{r^2} 
-\frac{(d-3)(d-4)kb}{r^2f}
-\frac{(d-1)(d-4)b}{ \ell^2f}
+\frac{(d-3)ba'}{ra}
\\
\nonumber
&&+\frac{(d-4)b'}{r}
-\frac{(d-4)kb'}{rf}
-\frac{(d-1)rb'}{\ell^2f}
+\frac{a'b'}{2a}
+\frac{b'^2}{b}~,
\end{eqnarray}
\begin{eqnarray}
\label{ep3} 
\frac{a'}{a}= 2\frac{ b\big[\ell^2(d-3)(d-4)(k-f)+(d-1)(d-2)r^2\big]
-(d-3)r\ell^2fb'}{r\ell^2f\big[rb'+2(d-3)b\big]}
\end{eqnarray}

\subsection{Asymptotics}
We consider solutions of the above equations whose boundary topology  
is the product of time and $S^{d-3}\times S^1$, $R^{d-3}\times S^1$ or $H^{d-3}\times 
S^1$. For even $d$, the solution of the Einstein equations
admits at large $r$  a power series expansion of the form:
\begin{eqnarray} 
\nonumber
a(r)&=&\frac{r^2}{\ell^2}+\sum_{j=0}^{(d-4)/2}a_j(\frac{\ell}{r})^{2j}
+c_z(\frac{\ell}{r})^{d-3}+O(1/r^{d-2}),
\\
\label{even-inf}
b(r)&=&\frac{r^2}{\ell^2}+\sum_{j=0}^{(d-4)/2}a_j(\frac{\ell}{r})^{2j}
+c_t(\frac{\ell}{r})^{d-3}+O(1/r^{d-2}),
\\
\nonumber
f(r)&=&\frac{r^2}{\ell^2}+\sum_{j=0}^{(d-4)/2}f_j(\frac{\ell}{r})^{2j}
+(c_z+c_t)(\frac{\ell}{r})^{d-3}+O(1/r^{d-2}),
\end{eqnarray}   
where $a_j,~f_j$ are constants depending on the index
$k$ and the spacetime dimension only. Specifically, we find
\begin{eqnarray}
\label{inf2}  
a_0=(\frac{d-4}{d-3})k~,~~
a_1=\frac{(d-4)^2k^2}{(d-2)(d-3)^2(d-5)},~~
a_2=-\frac{(d-4)^3(3d^2-23d+26)k^3}{3(d-2)^2(d-3)^3(d-5)(d-7)}~,
\end{eqnarray} 
\begin{eqnarray}
\label{inf3}  
f_0=\frac{k(d-1)(d-4)}{(d-2)(d-3)},~~
f_1=2a_1,~~
f_2=-\frac{2(d-4)^3(d^2-8d+11)k^3}{(d-2)^2(d-3)^3(d-5)(d-7)}~,
\end{eqnarray}   
their expression becoming more complicated for higher $j$, with no general
pattern becoming apparent.

The corresponding expansion for odd values of the spacetime 
dimension is given by:
\begin{eqnarray}
\nonumber 
a(r)&=&\frac{r^2}{\ell^2}+\sum_{j=0}^{(d-5)/2}a_j(\frac{\ell}{r})^{2j}
+\zeta\log(\frac {r}{\ell}) (\frac{\ell}{r})^{d-3}
+c_z(\frac{\ell}{r})^{d-3}+O(\frac{\log r}{r^{d-1}}),
\\
\label{odd-inf}
b(r)&=&\frac{r^2}{\ell^2}+\sum_{j=0}^{(d-5)/2}a_j(\frac{\ell}{r})^{2j}
+\zeta\log (\frac {r}{\ell}) (\frac{\ell}{r})^{d-3}
+c_t(\frac{\ell}{r})^{d-3}+O(\frac{\log r}{r^{d-1}}),
\\
\nonumber
f(r)&=&\frac{r^2}{\ell^2}+\sum_{j=0}^{(d-5)/2}f_j(\frac{\ell}{r})^{2j}
+2\zeta\log (\frac {r}{\ell}) (\frac{\ell}{r})^{d-3}
+(c_z+c_t+c_0)(\frac{\ell}{r})^{d-3}+O(\frac{\log r}{r^{d-1}}),
\end{eqnarray}   
where we note $\zeta=a_{(d-3)/2}\sum_{k>0}(d-2k-1)\delta_{d,2k+1}$,
while
\begin{eqnarray}
\label{inf4}
c_0=0~~~{\rm for }~~d=5, ~~
c_0=\frac{9k^3\ell^4}{1600}~~~{\rm for }~~d=7, 
~~
c_0=-\frac{90625 k^4\ell^6}{21337344}~~~{\rm for }~~d=9. 
\end{eqnarray} 

For any value of $d$, terms of higher order in $\frac{\ell}{r}$ depend only on the two 
constants $c_t$ and $c_z$. These constants are found numerically starting from the following expansion of the solutions near the event horizon (taken at constant $r=r_h$) and integrating the Einstein equations towards infinity: 
\begin{eqnarray}
\nonumber  
a(r)&=&
a_h
+\frac{2a_h(d-1)r_h}{(d-1)r_h^2+k(d-4)\ell^2}(r-r_h)
\\
\nonumber
&&+\frac{a_h(d-1)^2r_h^2}{\big[(d-1)r_h^2+k(d-4)\ell^2\big]^2}(r-r_h)^2 
+O(r-r_h)^3,
\\
\label{eh}
b(r)&=&b_1(r-r_h)
-\frac{b_1(d-4)\big[(d-1)r_h^2+(d-3)k\ell^2\big]}
{2(d-1)r_h^3+2(d-4)kr_h\ell^2}(r-r_h)^2
+O(r-r_h)^3,
\\
\nonumber
f(r)&=&\frac{1}{r_h \ell^2}\big[(d-1)r_h^2+k(d-4)\ell^2\big](r-r_h)
\\
\nonumber
&&-\frac{(d-4)}{2r_h^2\ell^2}
\big[(d-1)r_h^2+k(d-3)\ell^2\big](r-r_h)^2
+O(r-r_h)^3,
\end{eqnarray}
in terms of two parameters $a_h$, $b_1$. The condition for a regular 
event horizon is $f'(r_h)>0$, $b'(r_h)>0$. In the $k=-1$ case, this 
implies the existence of a minimal value of $r_h$, $i.e.$ for a given 
$\Lambda$:
\begin{eqnarray}
r_h>\ell\sqrt{(d-4)/(d-1)}.
\end{eqnarray}

Globally regular solutions with $r_h=0$ exist for $k=1$ only. The 
corresponding expansion near origin $r=0$ is:
\begin{eqnarray}
  \nonumber
a(r)&=&
\bar{a}_0
+\frac{\bar{a}_0(d-1)}{(d-2)}(\frac{r}{\ell})^2
+\frac{\bar{a}_0(d-1)^2}{d(d-2)^2(d-3)}(\frac{r}{\ell})^4+O(r^6),
\\
\label{reg1}
b(r)&=&
\bar{b}_0
+\frac{\bar{b}_0(d-1)}{(d-2)}(\frac{r}{\ell})^2
+\frac{\bar{b}_0(d-1)^2}{d(d-2)^2(d-3)}(\frac{r}{\ell})^4+O(r^6),
\\
\nonumber
f(r)&=&
1
+\frac{(d-1)(d-4)}{(d-2)(d-3)}(\frac{r}{\ell})^2
+\frac{2(d-1)^2}{d(d-2)^2(d-3)}(\frac{r}{\ell})^4+O(r^6),
\end{eqnarray}   
$\bar{a}_0$, $\bar{b}_0$ being positive constants.

\section{The properties of the solutions}
As with  the asymptotically flat case, one expects the values of 
mass and tension to be encoded in the constants $c_t$ and $c_z$ which 
appear in (\ref{even-inf}), (\ref{odd-inf}). However, in the presence 
of a non-vanishing cosmological constant, the generalization of 
Komar's formula is not straightforward and it requires the further 
subtraction of a contribution from the background configuration in order to 
render finite results when computing the conserved charges. 

While for 
$k=1$ one may subtract the globally regular configuration 
contribution, there is no obvious choice for such a background in the $k=-1$ 
case. Therefore we prefer to use a different approach and follow the 
general procedure proposed by Balasubramanian and Kraus 
\cite{Balasubramanian:1999re} to compute the conserved quantities for a spacetime 
with negative cosmological constant. This technique was inspired by the 
AdS/CFT correspondence and consists of adding to the action suitable 
boundary counterterms $I_{ct}$, which are functionals only of 
curvature invariants of the induced metric on the boundary. Such terms will 
not interfere with the equations of motion because they are intrinsic 
invariants of the boundary metric. By choosing appropriate 
counterterms, which cancel the divergences, one can then obtain well-defined 
expressions for the action and the energy momentum of the spacetime. 
Unlike the background subtraction methods, this procedure is intrinsic 
to the spacetime of interest and it is unambiguous once the 
counterterm action is specified.   

Thus we have to supplement the action in (\ref{action}) with 
\cite{Balasubramanian:1999re,Das:2000cu}:
\begin{eqnarray}
I_{\mathrm{ct}}^0 &=&\frac{1}{8\pi G}\int d^{d-1}x\sqrt{-\gamma 
}\left\{ -\frac{d-2}{\ell }-\frac{\ell \mathsf{\Theta }\left( d-4\right) 
}{2(d-3)}\mathsf{R}-\frac{\ell ^{3}\mathsf{\Theta }\left( d-6\right) 
}{2(d-3)^{2}(d-5)}\left(\mathsf{R}_{ab}\mathsf{R}^{ab}-\frac{d-1}{4(d-2)}\mathsf{R}^{2}\right) 
\right.
\nonumber  
\\
\label{Lagrangianct} 
&&+\frac{\ell ^{5}\mathsf{\Theta }\left( d-8\right) 
}{(d-3)^{3}(d-5)(d-7)}\left( 
\frac{3d-1}{4(d-2)}\mathsf{RR}^{ab}\mathsf{R}_{ab}-\frac{d^2-1}{16(d-2)^{2}}\mathsf{R}^{3}\right.  \nonumber \\
&&\left. -2\mathsf{R}^{ab}\mathsf{R}^{cd}\mathsf{R}_{acbd}\left. 
-\frac{d-1}{4(d-2)}\nabla _{a}\mathsf{R}\nabla ^{a}\mathsf{R}+\nabla 
^{c}\mathsf{R}^{ab}\nabla _{c}\mathsf{R}_{ab}\right) +...\right\} ,
\end{eqnarray}
where $\mathsf{R}$ and $\mathsf{R}^{ab}$ are the curvature and the 
Ricci tensor associated with the induced metric $\gamma $. The series 
truncates for any fixed dimension, with new terms entering at every 
new even value of $d$, as denoted by the step-function ($\mathsf{\Theta 
}\left( x\right) =1$ provided $x\geq 0$, and vanishes otherwise).

However, given the presence of $\log(r/\ell)$ terms in the asymptotic 
expansions (\ref{odd-inf}) (for odd $d$), the counterterms 
(\ref{Lagrangianct}) regularise the action for  even dimensions only. For odd 
values of $d$, we have to add the following extra terms to 
(\ref{action}) \cite{Skenderis:2000in}:
\begin{eqnarray}
I_{\mathrm{ct}}^{s} &=&\frac{1}{8\pi G}\int d^{d-1}x\sqrt{-\gamma 
}\log(\frac{r}{\ell})\left\{  
\mathsf{\delta }_{d,5}\frac{\ell^3 
}{8}(\frac{1}{3}\mathsf{R}^2-\mathsf{R}_{ab}\mathsf{R}^{ab}
)\right.
\nonumber  
\\
&&-\frac{\ell 
^{5}}{128}\left(\mathsf{RR}^{ab}\mathsf{R}_{ab}-\frac{3}{25}\mathsf{R}^{3} 
-2\mathsf{R}^{ab}\mathsf{R}^{cd}\mathsf{R}_{acbd}\left. -\frac{1}{10}\mathsf{R}^{ab}\nabla _{a}\nabla 
_{b}\mathsf{R}+\mathsf{R}^{ab}\Box \mathsf{R}_{ab}-\frac{1}{10}\mathsf{R}\Box 
\mathsf{R}\right)\delta_{d,7} +\dots
\right\}.\nonumber
\end{eqnarray}%
For the Kerr-AdS \cite{Gibbons:2004js} class of higher-dimensional 
rotating black holes in spaces with negative cosmological constant, these
terms vanish. However we shall see that they contribute non-trivially for the
solutions we obtain.

Using these counterterms in odd and even dimensions, one can 
construct a divergence-free boundary stress tensor from the total action 
$I=I_0+I_{\mathrm{ct}}^0+I_{\mathrm{ct}}^s$ by defining a boundary 
stress-tensor: 
\[
T_{ab}=\frac{2}{\sqrt{-\gamma}}\frac{\delta I}{\delta \gamma^{ab}}. 
\]
Thus a conserved charge 
\begin{equation}
{\frak Q}_{\xi }=\oint_{\Sigma }d^{d-2}S^{a}~\xi ^{b}T_{ab},
\label{Mcons}
\end{equation}%
can be associated with a closed surface $\Sigma $ (with normal 
$n^{a}$), provided the boundary geometry has an isometry generated by a 
Killing vector $\xi ^{a}$ \cite{Booth}. If $\xi =\partial /\partial 
t$ then $\mathfrak{Q}$ is the conserved mass/energy $M$. 
Similar to the $\Lambda=0$ case \cite{Harmark:2003dg}, there is also 
a second charge associated with  $\partial/\partial z$, corresponding 
to the solution's tension ${\mathcal T}$.

The computation of $T_{ab}$ is straightforward and we find the 
following expressions for mass and tension:  
\begin{eqnarray}
\label{MT} 
M&=&M_0+M_c^{(k,d)}~,~~M_0=\frac{\ell^{d-4}}{16\pi G 
}\big[c_z-(d-2)c_t\big]LV_{k,d-3}~,
\\
{\mathcal T}&=&{\mathcal T}_0+{\mathcal T}_c^{(k,d)}~,~~
{\mathcal T}_0=\frac{\ell^{d-4}}{16\pi G }\big[(d-2)c_z-c_t\big] 
V_{k,d-3}~,
\end{eqnarray}  
where $V_{k,d-3}$ is the total area of the angular sector. Here 
$M_c^{(k,d)}$ and ${\mathcal T}_c^{(k,d)}$ are  Casimir-like terms
which appear for an odd spacetime dimension only,
 \begin{eqnarray}
\label{MT-Cas} 
M_c^{(k,d)}=-L{\mathcal T}_c^{(k,d)} =\frac{\ell^{d-4}}{16\pi G 
}V_{k,d-3}L\left(\frac{1}{12}\delta_{d,5}
-\frac{333}{3200}\delta_{d,7}+\dots\right)~.
\end{eqnarray}  
We note that the considered Lorentzian solutions  extremize also the 
Euclidean action as the analytic continuation $t \to i\tau$ has no 
effect at the level of the equations of motion. The Hawking 
temperature of these solutions is computed by demanding regularity of the 
Euclideanized manifold as $r \to r_h$:
\begin{eqnarray}
T_H=\frac{1}{4\pi}\sqrt{\frac{b_1}{r_h\ell^2}\big[(d-1)r_h^2+k(d-4)\ell^2\big]}.
\label{Th}
\end{eqnarray}
Thus we can proceed further by formulating gravitational 
thermodynamics via the Euclidean path integral \cite{Hawking:ig}
\[
Z=\int D\left[ g\right] D\left[ \Psi \right] e^{-I\left[ g,\Psi 
\right]
}\simeq e^{-I}, 
\]%
Here, $D\left[g\right]$ is a measure on the space of metrics 
$g$, $D\left[\Psi\right]$ a measure on the 
space of matter fields $\Psi$, $I\left[ g,\Psi\right]$ is the 
action in terms of the metrics and matter fields and one 
integrates over all metrics and matter fields between some 
given initial and final Euclidean hypersurfaces, taking 
$\tau $ to have a period $\beta=1/T_H$. Semiclassically 
the result is given by the classical action evaluated on 
the equations of motion, and yields to 
this order an expression for the entropy: 
\begin{equation}
S=\beta M-I,  
\label{GibbsDuhem}
\end{equation}%
upon application of the quantum statistical relation to the partition 
function.
  
To evaluate the black string action, we integrate the Killing 
identity $\nabla^\mu\nabla_\nu  \zeta_\mu=R_{\nu \mu}\zeta^\mu,$
for the Killing vector $\zeta^\mu=\delta^\mu_t$, together with the  Einstein 
equation $R_t^t={(R - 2\Lambda)/2}$. Thus, we isolate the bulk action
contribution at infinity and at $r=0$ or $r=r_h$. The divergent 
contributions given by the surface integral term at infinity are also 
canceled by $I_{\rm{surface}}+I_{ct}$. Together with 
(\ref{GibbsDuhem}), we find $S=A_H/4G$, where $A_H=r_h^{d-3}V_{k,d-3}L\sqrt{a_h}$ is 
the event horizon area.
The same approach applied to the Killing vector $\zeta^\mu=\delta^\mu_z$ yields 
the result:
\begin{eqnarray}
\label{itot2}
I =-\beta {\mathcal T}L.
\end{eqnarray}
The relations (\ref{GibbsDuhem}) and (\ref{itot2}) lead to a simple 
Smarr-type formula, relating quantities defined at 
infinity to quantities defined at the event horizon:
\begin{eqnarray}
\label{smarrform} 
M+{\mathcal T}L=T_HS~,
\end{eqnarray}
(note that the corresponding relation in the $\Lambda=0$ case is 
$d-$dependent \cite{Harmark:2003dg}).

This relation also provides a useful check of the 
accuracy of the numerical solutions we obtain. We see now that in the limit of zero 
event horizon radius, the absolute values of the mass of solutions per
unit length of the extra-dimension $z$ and the tension are equal. 
\section{Numerical results}
Starting from the expansion (\ref{eh}) and using a standard ordinary  
differential  equation solver, we integrated the system 
(\ref{ep1})-(\ref{ep3}) adjusting  for fixed shooting parameters and integrating 
towards  $r\to\infty$. The integration stops when the asymptotic limit 
(\ref{even-inf}), (\ref{odd-inf}) is reached. Given 
$(k,~d,~\Lambda,~r_h)$, solutions with the right asymptotics exist for one set of  the 
shooting parameters  $(a_h,~b_1)$ only.

The results we present here are found for $\ell=1$. However, the 
solutions for any other values of the cosmological constant are easily 
found by using a suitable rescaling of the $\ell=1$ configurations. 
Indeed, to understand the dependence of the solutions on the 
cosmological constant, we note that the Einstein 
equations 
(\ref{ep1})-(\ref{ep3}) are left invariant by the transformation: 
\begin{eqnarray}
\label{transf1} 
r \to \bar{r}= \lambda r,~~\ell \to \bar{\ell}= \lambda \ell.
\end{eqnarray}

Therefore, starting from a solution corresponding to $\ell=1$ one may 
generate in this way a family of $\ell\neq 1$ vacuum solutions, which are termed 
``copies of solutions`` \cite{Harmark:2003eg}. The new solutions have the 
same length in the extra-dimension. Their relevant properties, 
expressed in terms of the corresponding properties of the initial solution, 
are as follows:
\begin{eqnarray}
\label{transf2} 
 \bar{r}_h=\lambda r_h,~\bar{\Lambda}=\Lambda/\lambda^2 ,~~
 \bar{T}_H=T_H/\lambda ,~~
 \bar{M}=\lambda^{d-4} M ,~~{\rm and}~~
 \bar{{\mathcal T}}=\lambda^{d-4} {\mathcal T}.
\end{eqnarray}
Now, given the full spectrum of solutions for $\ell=1$ (with 
$r_{min}<r_h<\infty$), one may find the corresponding branches for any value of 
$\Lambda<0$. 
Thus these solutions do not approach the uniform black 
string configurations 
as $\Lambda \to 0$ (note also that the local 
mass/tension of the $\Lambda=0$ black strings decay as $1/r^{d-4}$, 
whereas here the decay is $1/r^{d-3}$).

For $k=0$, the Einstein equations admit the exact solution $a=r^2$, 
$f=1/b=-2M/r^{d-3}+r^2/\ell^2$, which was recovered by our numerical 
procedure. This $k=0$ solution appears to be unique, corresponding to the 
known planar topological black hole, respectively to the AdS 
soliton 
\cite{AdSsoliton} (after an analytic continuation of the 
coordinates). Therefore in the remainder of this section we will concentrate on 
the $k=\pm 1$  cases only.

We have found  asymptotically AdS numerical solutions in all dimensions between five 
and twelve. We conjecture that they exist for any $d$ and, in the case $k=1$, for
any value of the event horizon $r_h$. For $d=5,~k=1$ our findings are in very 
good agreement with those presented in ref. \cite{Copsey:2006br}. 
%
\newpage
\setlength{\unitlength}{1cm}
\begin{picture}(18,7)
\centering
\put(2,0.0){\epsfig{file=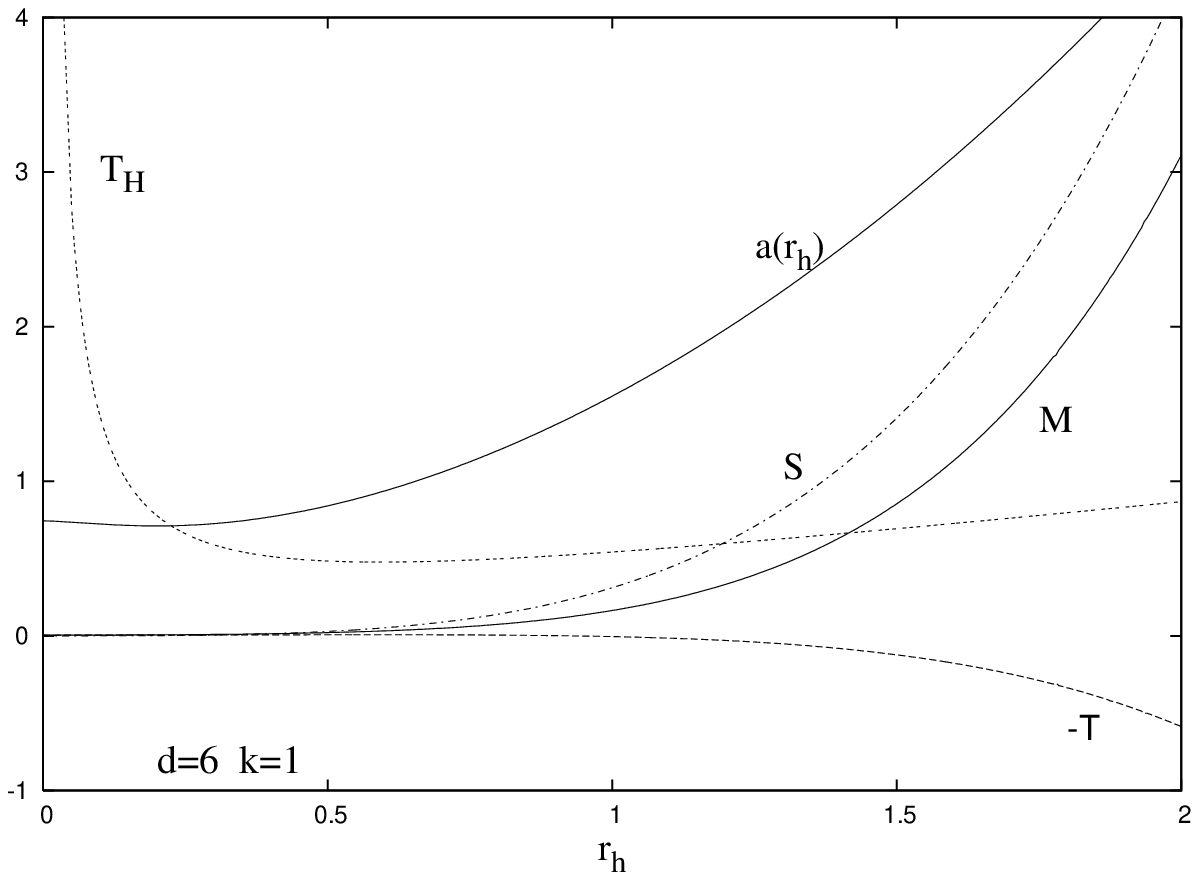,width=11cm}}
\end{picture}
\begin{picture}(19,8.)
\centering 
\put(2.4,0.0){\epsfig{file=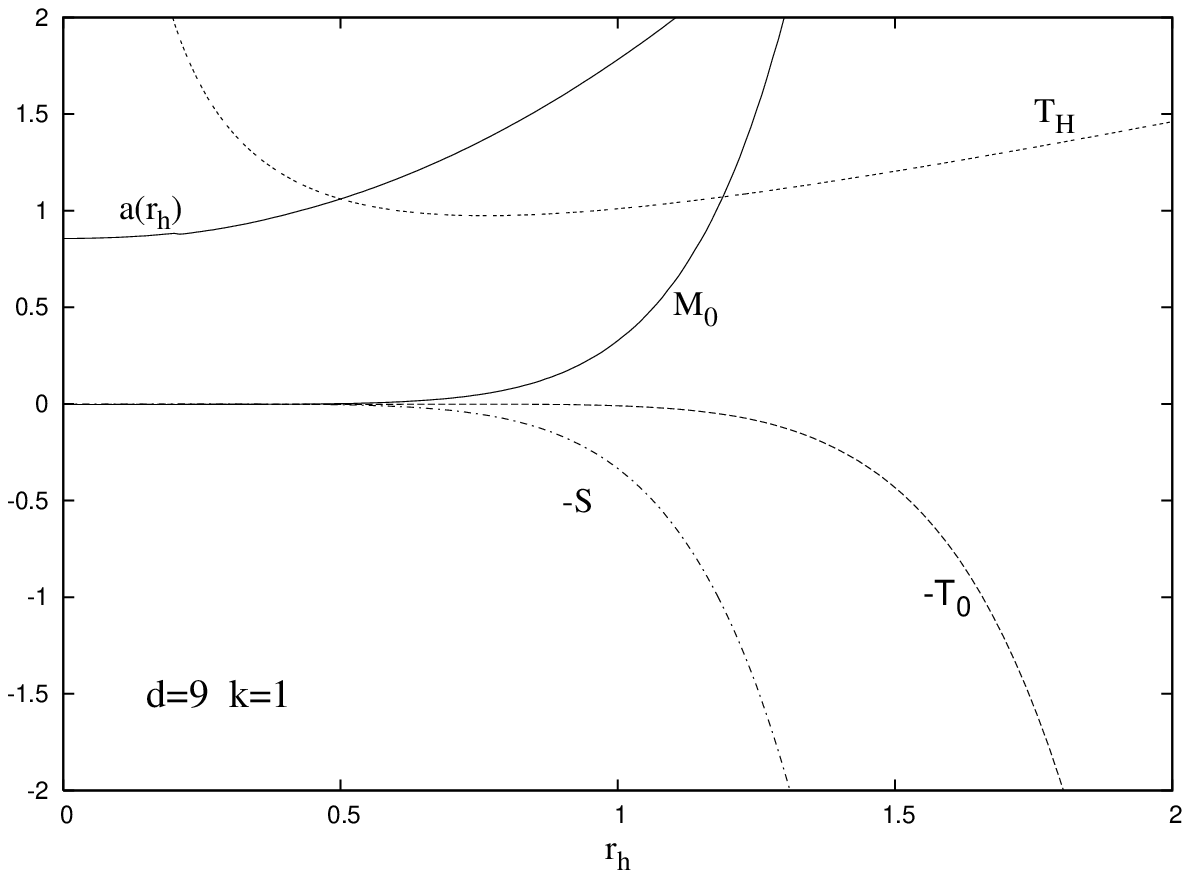,width=11.1cm}}
\end{picture}
\\
\\
{\small {\bf Figure 1.} The mass-parameter $M$, the tension 
${\mathcal T}$,  (without the Casimir terms) the value  of the metric function 
$a(r)$ at the event horizon as well as the Hawking temperature 
$T_H$ and the entropy $S$ of $k=1$ black string solutions 
are represented as functions of the event horizon radius in $d=6$, 
respectively $d=9$ dimensions. In the latter diagram we plot $-S,~-T_0$ so that
they can be easily distinguished from the curve for $M_0$.}
\\
\\
The $k=1$ solutions have a nontrivial zero event horizon radius limit 
corresponding to AdS vortices. As $r_h \to 0$ we find $e.g.$ 
$ c_t(d=6)\simeq -0.0801$, 
$ c_t(d=7)\simeq-0.0439$, 
$ c_t(d=8)\simeq 0.0403$,
$ c_t(d=9)\simeq 0.0229$,
 while 
$c_t(d=10)\simeq-0.0246$. 

%
 \newpage
 \setlength{\unitlength}{1cm}
 \begin{picture}(18,7)
 \centering
 \put(2,0.0){\epsfig{file=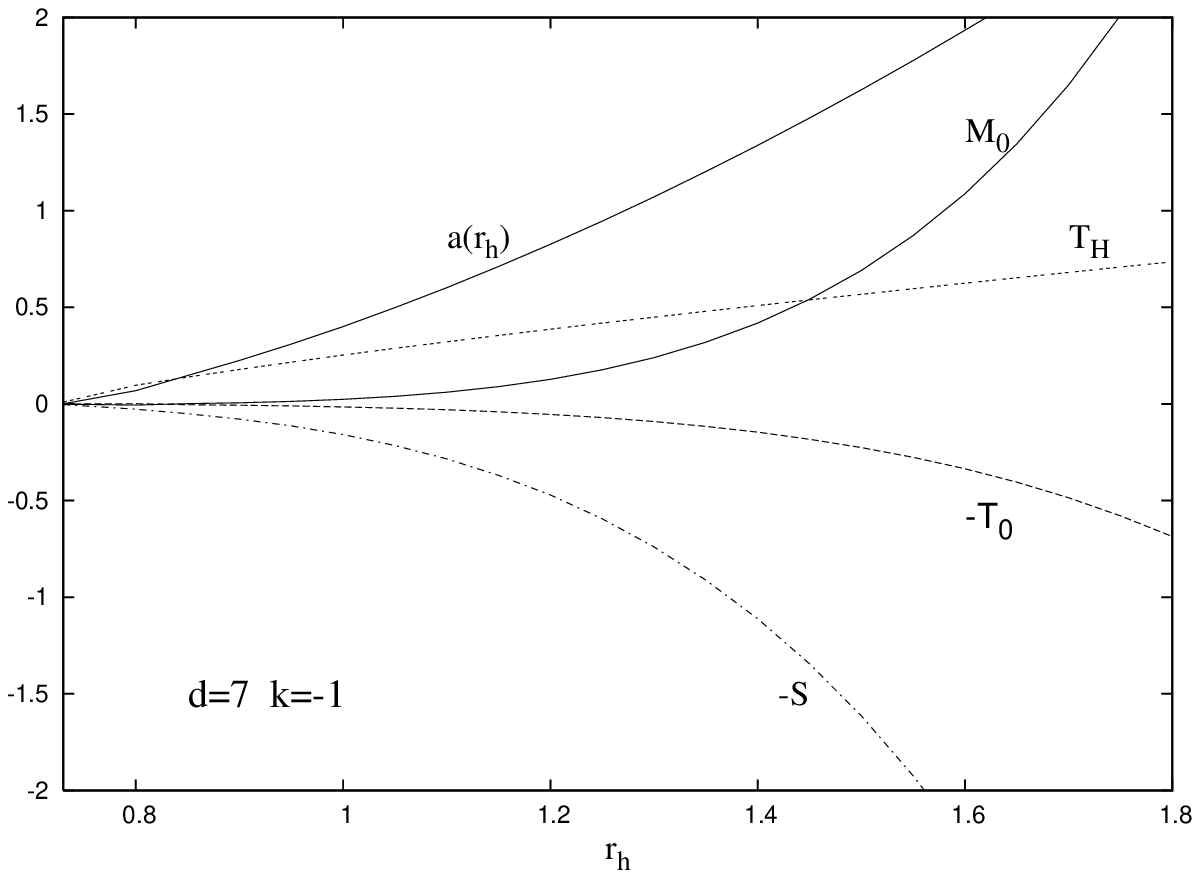,width=11cm}}
 \end{picture}
 \begin{picture}(19,8.)
 \centering 
 \put(2.4,0.0){\epsfig{file=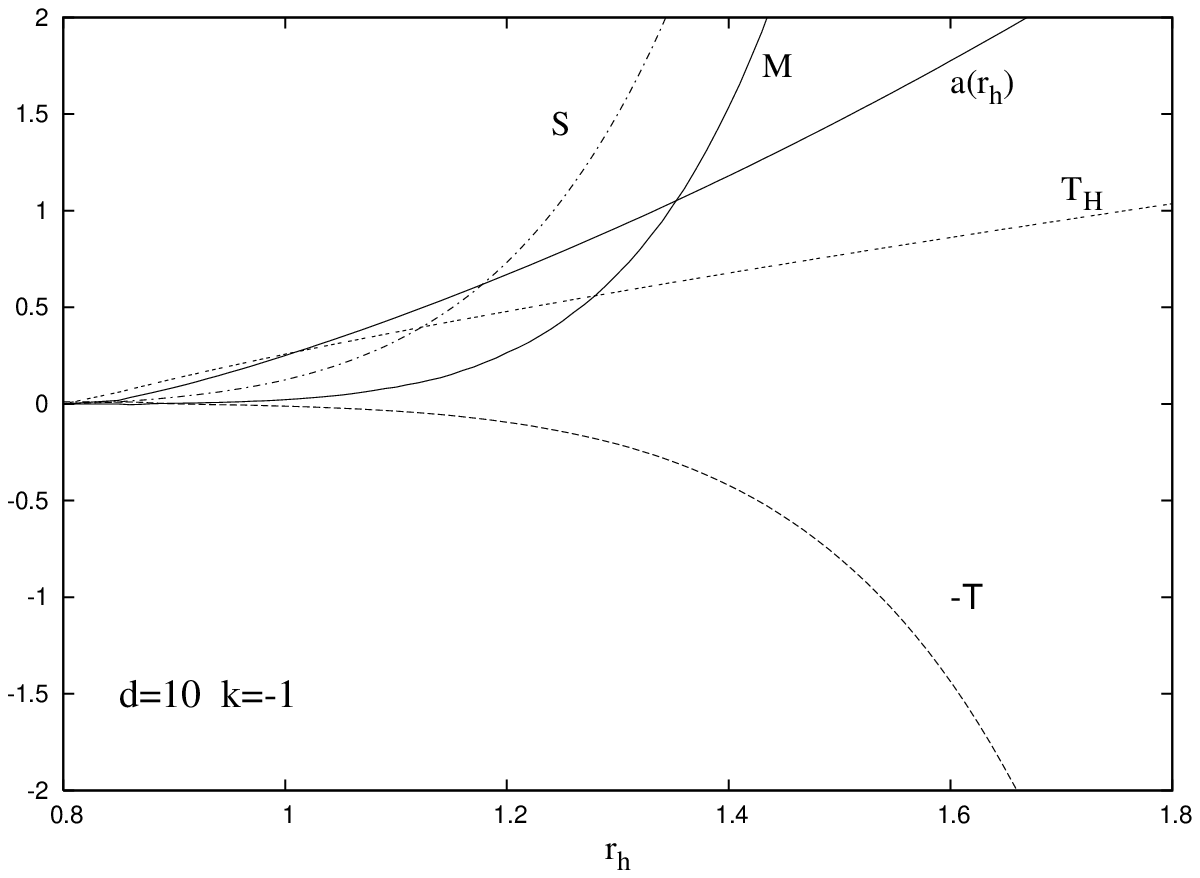,width=11.1cm}}
 \end{picture}
 \\
 \\
 {\small {\bf Figure 2.} Same as Figure 1 for $k=-1$
 black string solutions in dimensions $d=7$, respectively $d=10$.}
 \\
 \\
 In the same limit, we find $a(r)=b(r)$ with a nonzero value 
$a(0)=\bar a_0$, e.g. 
$\bar a_0(d=6)\simeq 0.744$, 
$\bar a_0(d=7)\simeq 0.797$, 
$\bar a_0(d=8)\simeq 0.8316$, 
$\bar a_0(d=9)\simeq 0.8561$, 
while 
$\bar a_0(d=10)\simeq 0.8743$\footnote{The appearance of these 
values  in the expansion at the origin suggests that analytic vortex 
solutions, if they exist, should be sought using a set of variables other
than ($a,~b,~f$).}. This assigns a nonzero mass and tension to 
the globally regular solutions according to (\ref{MT}).

The dependence of various physical parameters on the event horizon 
radius is presented in Figure 1 for $d=6$ and $d=9$ solutions with 
$k=1$ and Figure 2 for  $k=-1$ configurations in  $d=7$ and $d=10$ 
dimensions. 
\newpage
\setlength{\unitlength}{1cm}

\begin{picture}(18,7)
\centering
\put(2,0.0){\epsfig{file=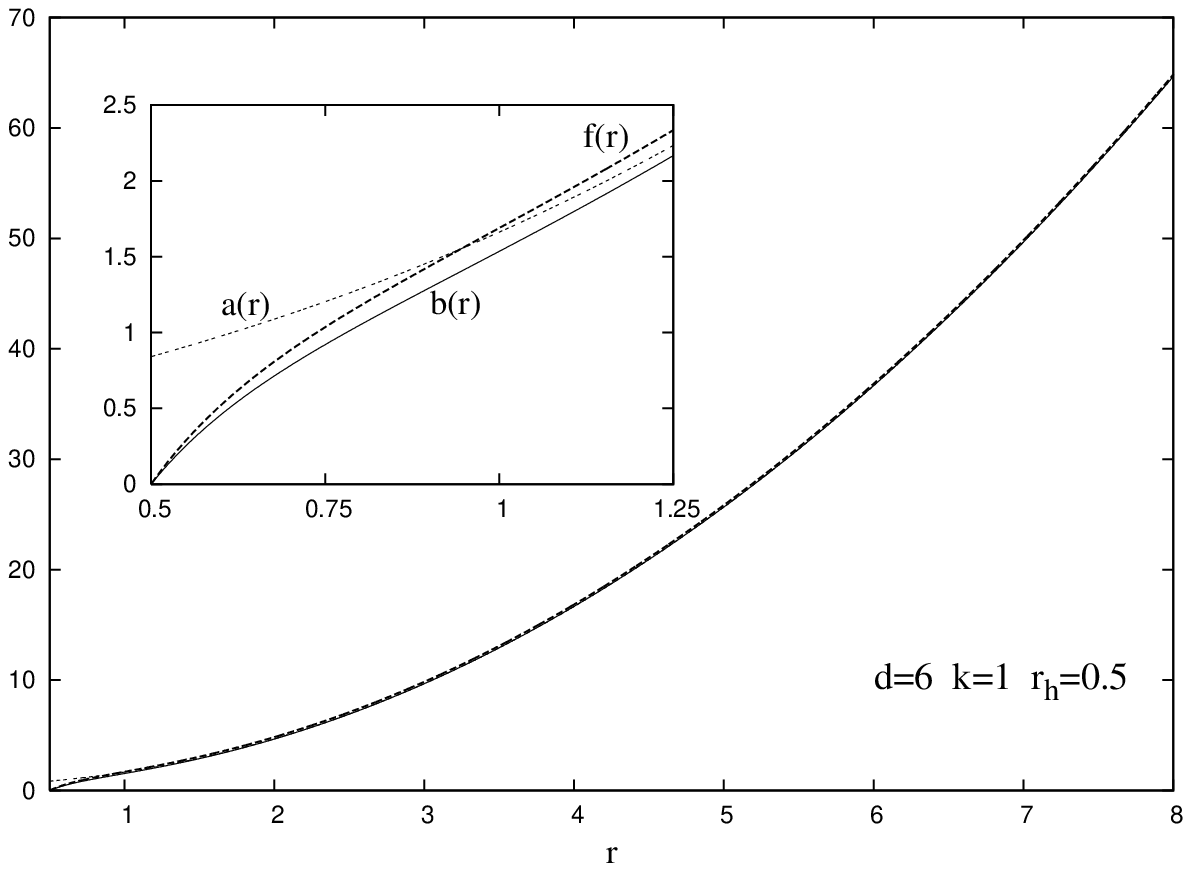,width=11cm}}
\end{picture}
{\small {\bf Figure 3.} The profiles of the metric functions $a(r)$, $b(r)$ and 
$f(r)$ are shown for a typical $k=1$ black string solution with $d=6$, 
$k=1$ and $r_h=0.5$.}
\\
\\
These plots retain the basic features of the solutions we found in other dimensions (note that we set $V_{k,d-3}L=1$ in the expressions for the mass, tension and entropy and we subtracted the Casimir 
energy and tension in odd dimensions). 

The mass, temperature, tension and entropy of $k=-1$ solutions increase 
monotonically with $r_h$.

For all the solutions we studied, the metric functions $a(r)$, $b(r)$ 
and $f(r)$ interpolate monotonically between the corresponding values 
at $r=r_h$ and the asymptotic values at infinity, without presenting 
any local extrema. As a typical example, in Figure $3$ the metric 
functions $a(r)$, $b(r)$ and $f(r)$  are
shown for a $d=6$, $k=1$ solution with  $r_h=0.5$, as functions of 
the radial coordinate $r$. One can see that the term $r^2/\ell^2$ 
starts dominating  the profile of these functions very rapidly, which implies 
a small difference between the metric functions for large enough $r$.

\section{$SL(2,R)$ symmetry in $(d-1)$-dimensions}
 Consider Einstein gravity coupled with a cosmological constant in $d$-dimensions. 
Its action is described by the Lagrangian:
\beqs
{\cal L}_d&=&\tilde{e}\tilde{R}-2\tilde{e}\Lambda,
\eeqs
where  $\tilde{e}=\sqrt{-\tilde{g}}$.

Let us assume that the fields are stationary and that the system admits two commuting Killing vectors (one of them is timelike, while the other corresponds to an isometry along a spatial direction $z$). We will perform a dimensional reduction from $d$-dimensions down to $(d-2)$-dimensions along the two directions $z$ and $t$. Our metric ansatz is:
\beqs
ds_{d}^2&=&e^{-\sqrt{\frac{2(d-3)}{d-2}}\phi}(dz+\chi dt)^2\nonumber\\
&&+e^{\sqrt{\frac{2}{(d-2)(d-3)}}\phi}\bigg[-e^{-\sqrt{\frac{2(d-4)}{d-3}}\phi_1}dt^2+e^{\sqrt{\frac{2}{(d-3)(d-4)}}\phi_1}ds_{d-2}^2\bigg].
\label{KKmetric}
\eeqs
In the first step of the dimensional reduction, reducing from $d$ to $(d-1)$-dimensions along the $z$-direction, one obtains the following $(d-1)$-dimensional Lagrangian\footnote{We denote $e=\sqrt{-g}$.}:
\beqs
{\cal L}_{d-1}&=&eR-2e\Lambda e^{\sqrt{\frac{2}{(d-2)(d-3)}}\phi}-\frac{1}{2}e(\partial\phi)^2-\frac{1}{4}ee^{-\sqrt{\frac{2(d-2)}{d-3}}\phi}({\cal F}_{(2)})^2,
\eeqs
while the metric and the matter fields in $(d-1)$-dimensions are given by the dilaton $\phi$, respectively the metric and the $1$-form potential:
\beqs
ds_{d-1}^2&=&-e^{-\sqrt{\frac{2(d-4)}{d-3}}\phi_1}dt^2+e^{\sqrt{\frac{2}{(d-3)(d-4)}}\phi_1}ds_{d-2}^2,\nonumber\\
{\cal A}_{(1)}&=&\chi dt.
\eeqs
Here ${\cal F}_{(2)}=d{\cal A}_{(1)}$ is the $2$-form field strength of the electromagnetic potential ${\cal A}_{(1)}$.

Performing a further dimensional reduction along the timelike direction, 
the Lagrangian in $(d-2)$-dimensions is given by:
\begin{eqnarray}
{\cal L}_{d-2}=eR
-2e\Lambda e^{\sqrt{\frac{2}{(d-2)(d-3)}}\phi
+\sqrt{\frac{2}{(d-3)(d-4)}}\phi_1}
-\frac{1}{2}e(\partial\phi)^2-\frac{1}{2}e(\partial\phi_1)^2
+\frac{1}{2}ee^{\sqrt{\frac{2(d-4)}{(d-3)}}\phi_1
-\sqrt{\frac{2(d-2)}{d-3}}\phi}(\partial\chi)^2
.\nonumber
\end{eqnarray}

We shall prove next that the above Lagrangian has a global symmetry group $SL(2,R)$. We perform first a rotation of the scalar fields:
\beqs
\left(\begin{array}{c}\hat{\phi}\\\hat{\phi_1}\end{array}\right)&=&\left(\begin{array}{cc}\sqrt{\frac{d-4}{2(d-3)}}&\sqrt{\frac{d-2}{2(d-3)}}\\\sqrt{\frac{d-2}{2(d-3)}}&-\sqrt{\frac{d-4}{2(d-3)}}\end{array}\right)\left(\begin{array}{c}\phi \\\phi_1\end{array}\right).
\label{rotsc}
\eeqs
Now, let us define the matrix ${\cal M}$ by:
\beqs {\cal M}&=&\left(\begin{array}{cc}e^{-\hat{\phi}_1}&\chi e^{-\hat{\phi}_1}\\\chi e^{-\hat{\phi}_1}&-e^{\hat{\phi}_1}
+\chi^2e^{-\hat{\phi}_1}\end{array}\right).\label{M}\eeqs
Note that $\det{\cal M}=-1$ hence ${\cal M}$ is not an $SL(2,R)$ matrix. Then it is easy to see that $(d-2)$-dimensional Lagrangian can be written in the following compact form:
\beqs
{\cal L}_{d-2}&=&eR-2e\Lambda e^{\frac{2\hat{\phi}}{\sqrt{(d-2)(d-4)}}}-\frac{1}{2}e(\partial\hat{\phi})^2+\frac{1}{4}e\tr[\partial{\cal 
M}^{-1}\partial{\cal M}].
\eeqs
The Lagrangian is manifestly invariant under $SL(2,R)$ transformations if one considers the following transformation laws for the potentials:
\beqs
g_{\mu\nu}&\ra& g_{\mu\nu},~~~~~~~ \Lambda\ra \Lambda,~~~~~~~{\cal M}\rightarrow ~\Omega^T{\cal M}\Omega,~~~~~~~\hat{\phi}\rightarrow \hat{\phi}.
\label{sltransform}
\eeqs
where $\Omega\in SL(2,R)$.

\section{Einstein-Maxwell-Liouville black holes}

Let us apply this technique to the new black string solutions in AdS backgrounds presented in the previous sections. Starting with the $d$-dimensional metric (\ref{metric}) and performing a double dimensional reduction along the $z$ and $t$ coordinates one can read directly the fields in the $(d-2)$-dimensional theory as:
\beqs
ds_{d-2}^2&=&\left(a b\right)^{\frac{1}{d-4}}\left(\frac{dr^2}{f}+r^2d\Sigma^2_{k,d-3}\right),\nonumber\\
e^{-\sqrt{\frac{2(d-3)}{d-2}}\phi}&=&a,~~~~~~~e^{-\sqrt{\frac{2(d-4)}{d-3}}\phi_1}=ba^{\frac{1}{d-3}},~~~~~~~\chi=0.
\eeqs
In order to apply the solution generating technique from the previous section, we shall perform the rotation of the scalar fields as given by (\ref{rotsc}). This yields
\beqs
e^{\hat{\phi}}&=&\left(a b\right)^{-\frac{1}{2}\sqrt{\frac{d-2}{d-4}}},~~~~~~~
e^{\hat{\phi}_1}=\left(\frac{b}{a}\right)^{\frac{1}{2}}.
\eeqs
We are now ready to apply the $SL(2,R)$-symmetry transformations as given in (\ref{sltransform}). For this purpose let us parameterize the matrix $\Omega$ in the form:
\beqs \Omega=\left(\begin{array}{cc} \alpha & \beta \\  \gamma & \delta \\\end{array}\right)
, \qquad \alpha\delta-\beta\gamma=1.
\eeqs
From the general form of the matrix ${\cal M}$ one can read the following scalar fields\footnote{In what follows we shall drop the prime symbol from the fields.}:
\beqs
e^{-\hat{\phi}_1'}&=&\frac{\alpha^2 a-\gamma^2 b}{\sqrt{a b}},~~~~~~~
\chi'=\frac{\alpha\beta a-\gamma\delta b}{\alpha^2 a-\gamma^2 b}.
\eeqs
Using now the inverse transformation of the one considered in (\ref{rotsc}) one finds the following scalar fields corresponding to the final solution:
\beqs
e^{\phi}&=&(\alpha^2 a-\gamma^2 b)^{-\sqrt{\frac{d-2}{2(d-3)}}},~~~~~~~
e^{\phi_1}=(a b)^{-\sqrt{\frac{d-3}{2(d-4)}}}(\alpha^2 a-\gamma^2 b)^{-\sqrt{\frac{d-4}{2(d-3)}}}.
\eeqs

Gathering all these results, one obtains in $(d-1)$-dimensions the following fields:
\beqs
ds_{d-1}^2&=&-ab(\alpha^2 a-\gamma^2 b)^{-\frac{d-4}{d-3}}dt^2+(\alpha^2 a-\gamma^2 b)^{\frac{1}{d-3}}\frac{dr^2}{f}+r^2(\alpha^2 a-\gamma^2 b)^{\frac{1}{d-3}}d\Sigma^2_{k,d-3},\nonumber\\
e^{\phi}&=&(\alpha^2 a-\gamma^2 b)^{-\sqrt{\frac{d-2}{2(d-3)}}},~~~~~~~
{\cal A}_{(1)}=\frac{\alpha\beta a-\gamma\delta b}{\alpha^2 a-\gamma^2 b}dt~.
\label{final}
\eeqs
which are a
solution of the equations of motion derived from the Lagrangian:
\beqs
{\cal L}_{d-1}&=&eR-2e\Lambda e^{\sqrt{\frac{2}{(d-2)(d-3)}}\phi}-\frac{1}{2}e(\partial\phi)^2-\frac{1}{4}ee^{-\sqrt{\frac{2(d-2)}{d-3}}\phi}({\cal F}_{(2)})^2,
\label{Ld-1}
\eeqs
which corresponds to an Einstein-Maxwell-Dilaton theory with a Liouville potential for the dilaton.

As a consistency check of our final solution, let us notice that if one takes $\Omega=I_2$ then one obtains the initial black string solution (\ref{metric}). Also, if $\alpha=\delta=\cosh p$ and $\beta=\gamma=\sinh p$ the effect of the $SL(2,R)$ transformation is equivalent to a boost of the initial black string solution in the $z$ direction.

In general, for a generic Kaluza-Klein dimensional reduction, if the isometry generated by the Killing vector $\frac{\partial}{\partial z}$ has fixed points then the dilaton $\phi$ will diverge and the $(d-1)$-dimensional metric will be singular at those points. However, this is not the case for our initial black string solutions and therefore the $(d-1)$-dimensional fields are non-singular in the near-horizon limit $r\ra r_h$. Indeed, in the near horizon limit $b(r)\ra 0$ while $a(r)\ra a_0$ and the $(d-1)$-dimensional 
fields are non-singular. The situation changes when we look in the asymptotic region. Recall from (\ref{KKmetric}) that 
\begin{eqnarray}
g_{zz}\equiv 
e^{-\sqrt{\frac{2(d-3)}{d-2)}}\phi}&=&(\alpha^2 a-\gamma^2 b)
\end{eqnarray}
gives the radius squared of the $z$-direction in $d$-dimensions and that it diverges in the large $r$ limit. Then for generic values of the parameters in $\Omega$ we find that $g_{zz}\sim r^2$ and the dilaton 
field in $(d-1)$-dimensions will diverge in the asymptotic region. Physically, this means that the spacetime decompactifies at infinity; the higher-dimensional theory should be used when describing such black holes in these regions. It is amusing to note that in this limit even though the Liouville potential goes rapidly to zero, the asymptotic structure of the $(d-1)$-dimensional metric is still non-standard. On the other hand one can choose the parameters in $\Omega$ such that $\alpha=\gamma$.
 In this case the  asymptotic behaviour of the $(d-1)$-dimensional fields is quite different as $g_{zz}\sim \frac{c_z-c_t}{r^{d-3}}$ and the radius of the $z$ direction collapses to zero asymptotically. However the dilaton field still diverges at infinity.

Note also that starting with the regular solution in $d$-dimensions  
one obtains a globally regular $(d-1)$-dimensional configuration upon 
applying the above solution-generating procedure. 
This is a solution of the Einstein-Dilaton system only, 
with a Liouville potential term for the dilaton. 
Indeed, starting with the near-origin expansion given in 
(\ref{reg1}) it is easy to see that the $(d-1)$-dimensional solution 
(\ref{final}) will be a globally regular solution of the 
Einstein-Dilaton system with a Liouville potential term of 
the dilaton once the condition $(\alpha^2-\gamma^2)a_0=1$ 
is satisfied. Notice that, since in the regular solution we have 
$a(r)=b(r)$, then the electromagnetic gauge field ${\cal A}_{(1)}$ 
is constant and, therefore, trivial. However, this is not the case for the dilaton. 

It is instructive to compare our solution of the equations of motion derived from (\ref{Ld-1}) 
with the previously known exact solutions from \cite{Chan} (see also \cite{Kosta} for the dyonic extensions). Consider for instance the solution given in eqs. ($6.8$) 
in \cite{Chan}:\footnote{ After canonically normalising the kinetic terms of the dilaton and the electromagnetic field, comparing the two actions one reads $n=d-1$, $a=\sqrt{d-2}$, $b=\frac{2}{(d-3)\sqrt{d-2}}$ so that $ab=\frac{2}{n-2}$. We also set $\phi_0=0$ for simplicity.}
\begin{eqnarray}
ds_{d-1}^2&=&-U(r)dt^2+\frac{dr^2}{U(r)}+\gamma^2r^{\frac{2}{d-1}}d\Omega_{d-3}^2,
\nonumber
\\
U(r)&=&r^{\frac{2(d-2)}{d-1}}\left(-\frac{(d-1)^2(d-4)}{2\gamma^2(d-3)^2}-
\frac{4M(d-1)}{(d-3)\gamma^{d-3}r^{\frac{2(d-3)}{d-1}}}+\frac{Q^2(d-1)^2}
{(d-3)^2\gamma^{2(d-3)}}\frac{1}{r^{\frac{4(d-2)}{d-1}}}\right),
\nonumber
\\
e^{\sqrt{\frac{2}{(d-2)(d-3)}}\phi}&=&r^{-\frac{2}{d-1}},~~~~~~~
A_t=\frac{(d-1)Q}{(d-3)\gamma^{d-3}}\frac{1}{r^{\frac{2(d-3)}{d-1}}},~~~~~~~
\Lambda=\frac{(d-2)(d-4)}{2\gamma^2}.\nonumber
\end{eqnarray}
Lifting now the solution to $d$-dimensions by using (\ref{KKmetric}), performing the coordinate transformation $R=\left(\frac{d-1}{d-3}\right)r^{\frac{d-3}{d-1}}$ and rescaling the $z$-coordinate to absorb a constant factor one obtains:
\beqs
ds_{d}^2&=&R^2\left(dz-\frac{J}{2R^2}dt\right)^2-\left(-\tilde{M}-\frac{R^2}{l^2}+\frac{J^2}{4R^2}\right)dt^2+\left(-\tilde{M}-\frac{R^2}{l^2}+\frac{J^2}{4R^2}\right)^{-1}dR^2+\gamma^2d\Omega_{d-3}^2,\nonumber
\eeqs
where we defined:
\beqs
\nonumber
J&=&\frac{2Q(d-1)^2}{(d-3)^2\gamma^{d-3}},~~~~~\tilde{M}=\frac{4(d-1)M}{(d-3)\gamma^{d-3}},~~~~~\Lambda=\frac{(d-2)}{l^2}.
\eeqs
It is now manifest that the $d$-dimensional solution is the direct product of the analytic continuation of the three-dimensional rotating BTZ black hole \cite{BTZ} with a $(d-3)$-dimensional sphere. Note that if one analytically continues the parameters $\gamma$, $l$  and exchanges the sphere with the metric of the hyperboloid $H^{d-3}$, one obtains by dimensional reduction a non-asymptotically flat black hole with hyperbolic topology. By contrast, in our $d$-dimensional black string solutions there is a non-
trivial warp factor multiplying the metric element of the sphere/hyperboloid; therefore the $(d-1)$-dimensional black hole solution is clearly different. Similar topological black hole solutions of Einstein-Maxwell-Dilaton theory with a Liouville potential for the dilaton that arise via dimensional reduction from higher dimensions have been discussed in \cite{Cai:1996eg, Yazadjiev:2005du}.

\section{Discussion}
In this work we have presented arguments for the existence of a new 
class of solutions of Einstein gravity with negative cosmological 
constant. For such solutions the topological structure of the boundary is 
the product of time and  $S^{d-3}\times R$  or 
$H^{d-3}\times R$  and they correspond to 
black strings with the horizon topology $S^{d-3}\times S^1$ or $H^{d-3}\times R$ respectively
(here the black string is wrapping the $S^1$ circle). More generally, we can replace the $(d-3)$-dimensional sphere (hyperboloid) with any Einstein space with positive (negative) curvature, normalised such that its Ricci tensor is $R_{ab}=k(d-4)g_{ab}$, with $k=0, \pm 1$.

We expect these solutions to be relevant in the context of AdS/CFT and more generally in the context of gauge/gravity dualities. Let us consider for example the $5$-dimensional black string solution. The background metric upon which the dual field theory resides is found by taking the rescaling 
$h_{\mu \nu}=\lim_{r \rightarrow \infty} \frac{\ell^2}{r^2}\gamma_{\mu \nu}$.

Restricting to the five-dimensional 
$k=\pm 1$ black strings, 
 we find
 \begin{eqnarray}
ds^2=h_{ab}dx^adx^b=-dt^2+dz^2+\ell^2d\Sigma_k^2,
\end{eqnarray}
and so the conformal boundary, where the $\mathcal{N}=4$ SYM theory lives, 
is $R\times S^1\times S^2$ for $k=1$ or $R\times S^1\times H^2$ 
for $k=-1$.

The expectation value of the stress tensor of the dual CFT 
can be computed using the  relation 
\cite{Myers:1999ps}:
\begin{eqnarray} 
\sqrt{-h}h^{ab}<\tau _{bc}>=\lim_{r\rightarrow \infty }\sqrt{-\gamma 
}\gamma
^{ab}T_{bc},
\end{eqnarray}
which gives
\begin{eqnarray}  \label{st}
&&<\tau^t_t>=\frac{36c_t-12c_z-1}{192\pi G\ell} , ~~<\tau^z_z>=\frac{-12c_t+36c_z-1}{192\pi G\ell}, 
\\
\nonumber
&&<\tau^{\theta}_{\theta}>=<\tau^{\phi}_{\phi}>=\frac{-12(c_t+c_z)+2}{192\pi G\ell}.
\end{eqnarray}
As expected, this stress tensor  is finite and covariantly 
conserved. However it is $not$ traceless and its trace is precisely 
equal to the conformal anomaly of the boundary CFT 
\cite{Skenderis:2000in}:
\beqs
{\cal A}=-\frac{2\ell^3}{16\pi G}\left(-\frac{1}{8}
\mathsf{R}_{ab}\mathsf{R}^{ab}+\frac{1}{24}\mathsf{R}^2\right).
\eeqs
A similar computation performed for the seven-dimensional case leads 
to a boundary stress tensor whose trace matches precisely the 
conformal anomaly of the dual six-dimensional superconformal $(2,0)$ theory 
\cite{Skenderis:2000in,Bastianelli:2000hi}.
 
We note that the analytic continuation $z \to i u $, $t \to i\chi$ 
in the general line element (\ref{metric}) gives a bubble solution:
\begin{eqnarray}
\label{metric-b} 
ds^2=-a(r)du^2+b(r)d\chi^2+ \frac{dr^2}{f(r)}
+r^2d\Sigma^2_{k,d-3},
\end{eqnarray}
(where $\chi$ has a periodicity $\beta=1/T_H$) whose properties
can be discussed by using the methods in \cite{Copsey:2006br}. We find for instance a `small' bubble that is the analytic continuation of the small black hole solution and also a `large' bubble, which is the analytical continuation of the large black hole solution. Note that since we found that $a(r)=b(r)$ for our regular solution, its analytic continuation leads to the same regular solution. Using the counterterm approach it is possible to compute the mass of these bubble solutions with the result that:
\beqs
M_{bubble}&=&-\beta T.
\eeqs
\newpage
\setlength{\unitlength}{1cm}

\begin{picture}(18,7)
\centering
\put(2,0.0){\epsfig{file=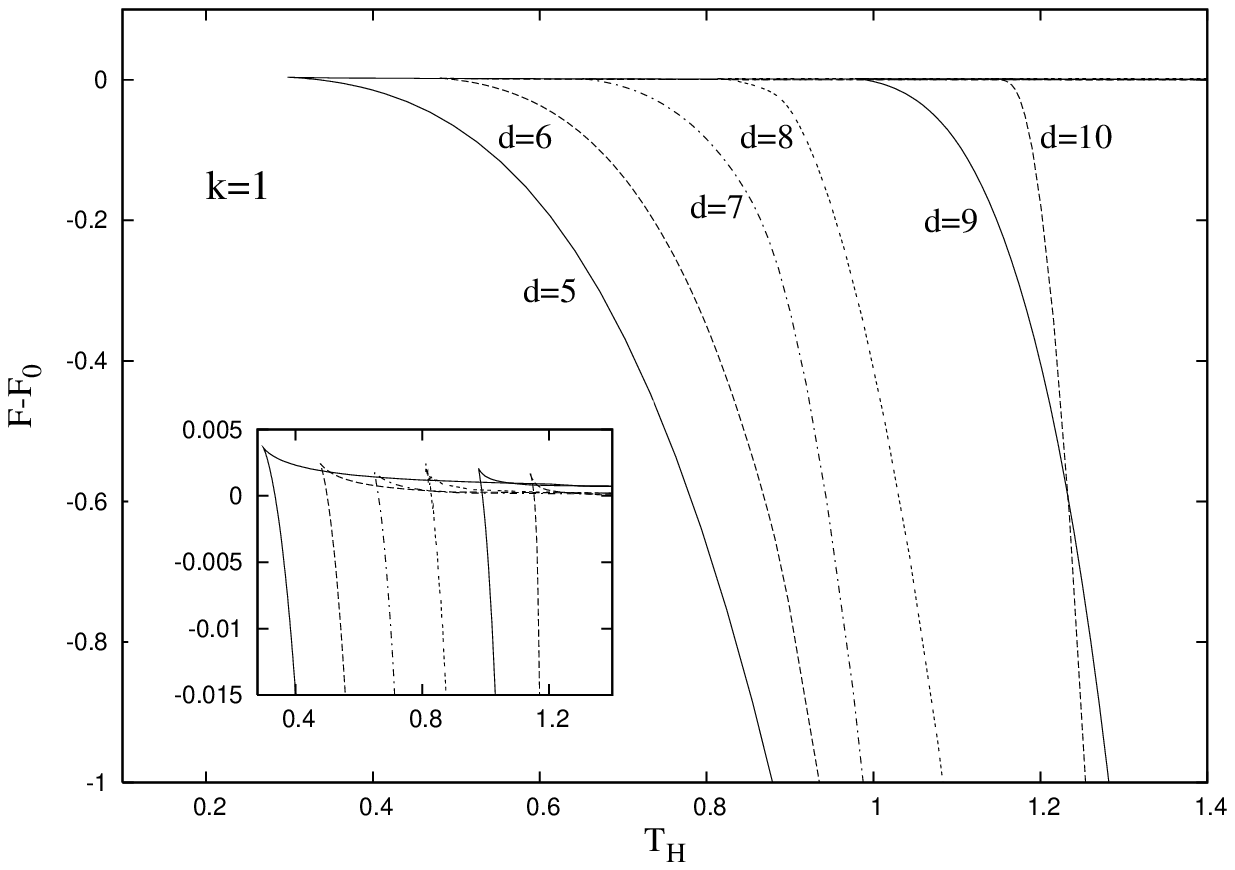,width=11cm}}
\end{picture}
{\small {\bf Figure 4.} The free energy \textit{vs.} the temperature for the small 
and large $k=1$ black string solutions. Here we have subtracted the free 
energy contribution $F_0$ of the globally regular solution. Note that the branch corresponding to the small black holes is always positive, while the one for large black holes changes sign, rapidly becoming negative at high temperatures.}
\\
\\
where $T$ is given in (\ref{MT}) while $\beta=1/T_H$, where $T_H$ is the Hawking temperature given in (\ref{Th}). 
We find then that for small values of $\beta$ (\textit{i.e.} small size of the $\chi$-circle at infinity) the `large' bubble solution has less mass than either the regular solution or the `small' bubble \footnote{We thank Keith Copsey and Gary Horowitz for pointing this out to us.}, while for circles with size large enough the background solution has the minimum energy.
 
As with  the spherically symmetric Schwarzschild-AdS solutions, the temperature of the $k=1$ 
black string solutions is bounded from below, as we can see in Figures $1$ and $2$. At low temperatures we have a single bulk solution, which we conjecture to correspond to the 
thermal globally regular solution. At high temperatures there exist two 
additional solutions that correspond to the small and large black holes. 
The free energy $F=I/\beta$ of the $k=1$ solutions is positive for small $r_h$ and negative for large $r_h$. This suggests that the phase transition found in \cite{Hawking:1982dh} occurs here as well.  Indeed, in Figure $4$ we plot the free energy versus the temperature for the small and large black hole solutions for $5\leq d\leq 10$. We observe the physics familiar from the Schwarzschild-AdS case: we have the two 
branches consisting of smaller (unstable) and large (stable) black holes. The entire unstable branch has positive free energy while the stable branch' free energy goes rapidly negative for all $T >T_c$. Here 
$T_c$ is the critical temperature at which we observe a phase transition between the large black holes and the thermal globally regular background.
  
As avenues for further research, it would be interesting to consider black hole solutions with an 
$S^{d-2}$ event horizon topology, whose conformal infinity is the product 
of time and $S^{d-3}\times S^1$,  presenting the asymptotic expansion (\ref{even-inf}), (\ref{odd-inf}). 
Such solutions are known to exist for $\Lambda=0$, approaching asymptotically  $(d-1)$-dimensional Minkowski space times a circle ${\cal M}^{d-1}\times S^1$ \cite{Kudoh:2004hs}. Also, the black string solutions may be useful in finding the AdS version of the $d=5$ asymptotically flat black rings \cite{Emparan:2001wn}. The heuristic construction of  black rings discussed in \cite{Emparan:2004wy} applies in this case too, and we expect an AdS black ring to approach the black string solution in the limit where the radius of the ring circle grows very large.  

Another interesting issue to investigate is the Gregory-Laflamme instability \cite{Gregory:1993vy}. We expect the black string solutions discussed in this paper to be unstable for some critical values of the parameters. 

We also note that the existence of globally regular solutions and of black hole solutions suggests that there might be some kind of critical phenomena associated with the collapse of matter to the black string, in keeping with the critical phenomena encountered in the study of the gravitational collapse of matter fields in spaces with spherical symmetry (see \cite{Gundlach:2002sx} and the references therein).  This is because a given distribution of matter undergoing collapse could potentially form either of these solutions, depending upon certain parameters in the initial data.  At the bifurcation point one presumably would see critical phenomena. A consideration of these aspects is outside the scope of this work. 

Further analysis of these metrics and their role in string theory remain interesting issues to explore in the future.

\medskip

\section*{Acknowledgements}
 The work of E.R. was carried out in the framework of Enterprise--Ireland 
 Basic Science Research Project SC/2003/390. The work of R.B.M. and C.S. was supported by the Natural Sciences \& Engineering Research Council of Canada.

\end{document}